\documentclass[prb,twocolumn]{revtex4}

\usepackage{amsmath}
\usepackage{amssymb}
\usepackage{amsfonts}
\usepackage{graphicx}
\usepackage{subfigure}
\usepackage{wrapfig}
\usepackage{float}
\floatstyle{boxed}
\newfloat{program}{h}{lop}
\floatname{program}{}

\begin{document}

\title{MATLAB codes for teaching quantum physics: Part 1}

\author{R.~Garcia}
\email{garcia@wpi.edu}

\author{A. Zozulya}

\author{J. Stickney}

\affiliation{Department of Physics, Worcester Polytechnic Institute,
Worcester, MA 01609}

\date{\today}

\begin{abstract}
Among the ideas to be conveyed to students in an introductory
quantum course, we have the pivotal idea championed by Dirac that
functions correspond to column vectors (kets) and that differential
operators correspond to matrices (ket-bras) acting on those vectors.
The MATLAB (matrix-laboratory) programming environment is especially 
useful in conveying these concepts to students because it
is geared towards the type of matrix manipulations useful in 
solving introductory quantum  physics problems. In this article, we 
share MATLAB codes which have been developed at WPI, focusing on 1D 
problems, to be used in conjunction with Griffiths' introductory text.
\end{abstract}

\maketitle

Two key concepts underpinning quantum physics are the Schrodinger
equation and the Born probability equation. In 1930 Dirac introduced
bra-ket notation for state vectors and operators.\cite{dirac} This 
notation emphasized and clarified the role of inner products and linear
function spaces in these two equations and is fundamental to our
modern understanding of quantum mechanics. The Schrodinger
equation tells us how the state $\Psi$ of a particle evolves in
time. In bra-ket notation, it reads
\begin{equation}
i\hbar \frac {d}{dt}|\Psi \rangle = {H} |\Psi \rangle \label{eq:one}
\end{equation}
where ${H}$ is the Hamiltonian operator and $|\Psi \rangle$ is a ket or
column vector representing the quantum state of the particle. When 
a measurement of a physical quantity $A$ is made on a particle 
initially in the state $\Psi$, the Born equation provides a way 
to calculate the probability $P(A_o)$ that a particular result 
$A_o$ is obtained from the measurement. In bra-ket notation, it reads
\cite{proportionality} 
\begin{equation}
P(A_o ) \thicksim | \langle A_o|\Psi \rangle|^{2} \label{eq:two}
\end{equation}
where if $|A_o\rangle$ is the state vector corresponding to the particular
result $A_o$ having been measured, $\langle A_o|=|A_o \rangle^\dagger$ is the
corresponding bra or row vector and $\langle A_o|\Psi \rangle$ is thus the
inner product between $|A_o\rangle$ and $|\Psi\rangle$.
In the Dirac formalism, the correspondence between the wavefunction
$\Psi(\vec{x})$ and the ket $|\Psi \rangle$ is set by the relation
$\Psi(\vec{x})=\langle\vec{x}|\Psi\rangle$, where $|\vec{x}\rangle$
is the state vector corresponding to the particle being located at
$\vec{x}$. Thus we regard $\Psi (\vec{x})$ as a component of a state
vector $|\Psi\rangle $, just as we usually \cite{gibbsheaviside} regard 
$a_{i} = $ \^i$ \cdot \vec{a}$ as a component of $\vec{a}$ along the 
direction \^i.  Similarly, we think of the Hamiltonian operator as a
matrix
\begin{equation}
{H} = \int d^3\vec{x}~ |\vec{x}\rangle \left\lbrace \frac{1}{2m}\left( \frac{%
\hbar }{i}\frac{\partial }{\partial \vec{x}}\right) ^{2}\Psi (\vec{x},t)+U(%
\vec{x}) \right\rbrace \langle \vec{x}| \label{eq:three}
\end{equation}
acting on the space of kets.

While an expert will necessarily regard Eqs.(1-3) as a great
simplification when thinking of the content of quantum physics, the
novice often understandably reels under the weight of the immense
abstraction. We learn much about student thinking from from the
answers given by our best students. For example, we find a common 
error when studying 1D quantum mechanics is a student treating 
$\Psi(x)$ and $|\Psi\rangle$ interchangeably, ignoring the fact that 
the first is a scalar but the ket corresponds
to a column vector. For example, they may write incorrectly
\begin{equation}
\langle p| |\Psi\rangle |x\rangle = |\Psi\rangle \langle p|x \rangle
\mbox{ \hspace{4mm} (incorrect!)}
\end{equation}
or some similar abberation. To avoid these types of misconceptions, 
a number of educators and textbook authors have stressed incorporating 
a numerical calculation aspect to quantum 
courses.\cite{rob,gould,spenser,styer,goldberg,singh} The motive is simple. 
Anyone who has done numerical calculations can't help but regard
a ket $|\Psi\rangle$ as a column vector, a bra $\langle \Psi|$ as a
row vector and an operator $H$ as a matrix because that is how they
concretely represented in the computer. Introducing a computational
aspect to the course provides one further benefit: it gives the
beginning quantum student the sense that he or she is being
empowered to solve real problems that may not have simple, analytic
solutions.

With these motivations in mind, we have developed MATLAB 
codes \cite{matlabwebsite} for solving typical 1 D problems found 
in the first part of a junior level quantum course based on Griffith's 
book.\cite{griffiths} We chose MATLAB for our programming 
environment because the MATLAB syntax is especially simple for 
the typical matrix operations used in 1D quantum mechanics problems 
and because of the ease of plotting functions. While some MATLAB 
numerical recipes have previously been published by others,\cite{garcia,lindblad}
the exercises we share here are special because they emphasize simplicity 
and quantum pedagogy, not numerical efficiency. Our point has been 
to provide exercises which show students how to numerically solve 
1 D problems in such a way that emphasizes the column vector aspect 
of kets, the row vector aspect of bras and the matrix aspect of 
operators. Exercises using more efficient MATLAB ODE solvers or 
finite-element techniques are omitted because they do not serve this 
immediate purpose. In part II of this article, we hope to share
MATLAB codes which can be used in conjunction with teaching topics
pertaining to angular momentum and non-commuting observables.

\section{Functions as Vectors}

To start students thinking of functions as column vector-like
objects, it is very useful to introduce them to plotting and integrating
functions in the MATLAB environment. Interestingly enough, 
the plot command in MATLAB takes vectors as its basic input element. 
As shown in Program 1 below, to plot a function $f(x)$ in MATLAB, we 
first generate two vectors: a vector of $x$ values and a vector of $y$ 
values where $y=f(x)$. The command $plot(x,y,'r')$ then generates a 
plot window containing the points $(x_i,y_i)$ displayed as red 
points ($'r'$). Having specified both $x$ and $y$, to evaluate the 
definite integral $\int_{-L}^L y dx$, we need only sum all the $y$ 
values and multiply by $dx$.

\begin{program}
\scriptsize
\begin{verbatim}
%****************************************************************
% Program 1: Numerical Integration and Plotting using MATLAB
%****************************************************************
N=1000000;               % No. of points
L=500;                   % Range of x: from -L to L
x=linspace(-L,L,N)';     % Generate column vector with N
                         % x values ranging from -L to L
dx=x(2)-x(1);            % Distance between adjacent points

% Alternative Trial functions:
% To select one, take out the comment command % at the beginning.
%y=exp(-x.^2/16);                 % Gaussian centered at x=0
%y=((2/pi)^0.5)*exp(-2*(x-1).^2); % Normed Gaussian at x=1
%y=(1-x.^2).^-1;            % Symmetric fcn which blows up at x=1
%y=(cos(pi*x)).^2;          % Cosine fcn
%y=exp(i*pi*x);             % Complex exponential
%y=sin(pi*x/L).*cos(pi*x/L);% Product of sinx times cosx
%y=sin(pi*x/L).*sin(pi*x/L);% Product of sin(nx) times sin(mx)
%A=100; y=sin(x*A)./(pi*x); % Rep. of delta fcn
A=20; y=(sin(x*A).^2)./(pi*(A*x.^2));% Another rep. of delta fcn

% Observe: numerically a function y(x) is represented by a vector!

% Plot a vector/function
plot(x,y);                             % Plots vector y vs. x
%plot(x,real(y),'r', x, imag(y), 'b'); % Plots real&imag y vs. x
axis([-2 2 0 7]);    % Optimized axis parameters for sinx^2/pix^2
%axis([-2 2 -8 40]); % Optimized axis parameters for sinx/pix

% Numerical Integration
sum(y)*dx        % Simple numerical integral of y
trapz(y)*dx      % Integration using trapezoidal rule
\end{verbatim}
\end{program}

\section{Differential Operators as Matrices}

Just as $f(x)$ is represented by a column vector $|f\rangle$ 
in the computer, for numerical purposes a differential operator 
$\hat{D} $ acting on $f(x)$ is reresented by a matrix $D$ that 
acts on $|f\rangle$. As illustrated in Program 2, MATLAB provides 
many useful, intuitive, well-documented commands for generating 
matrices $D$ that correspond to a given $\hat{D}$.\cite{matlabwebsite} 
Two examples are the commands $ones$  and $diag$. The command 
$ones(a,b)$ generates an $a \times b$ matrix of ones. The 
command $diag(A,n)$  generates a matrix with the elements of 
the vector $A$ placed along the $n^{th}$ diagonal and zeros 
everywhere else. The central diagonal corresponds to $n=0$, the 
diagonal above the center one corresponds to $n=1$, etc...). 

An exercise we suggest is for students to verify that the derivative 
matrix is not Hermitian while the derivative matrix
times the imaginary number $i$ is. This can be very valuable for
promoting student understanding if done in conjunction
with the proof usually given for the differential operator.

\begin{program}
\scriptsize
\begin{verbatim}
%****************************************************************
% Program 2: Calculate first and second derivative numerically
% showing how to write differential operator as a matrix
%****************************************************************
% Parameters for solving problem in the interval 0 < x < L
L = 2*pi;                   % Interval Length
N = 100;                    % No. of coordinate points
x = linspace(0,L,N)';       % Coordinate vector
dx = x(2) - x(1);           % Coordinate step

% Two-point finite-difference representation of Derivative
D=(diag(ones((N-1),1),1)-diag(ones((N-1),1),-1))/(2*dx);
% Next modify D so that it is consistent with f(0) = f(L) = 0
D(1,1) = 0; D(1,2) = 0; D(2,1) = 0;	% So that f(0) = 0
D(N,N-1) = 0; D(N-1,N) = 0; D(N,N) = 0; % So that f(L) = 0

% Three-point finite-difference representation of Laplacian
Lap = (-2*diag(ones(N,1),0) + diag(ones((N-1),1),1) ...
                           + diag(ones((N-1),1),-1))/(dx^2);
% Next modify Lap so that it is consistent with f(0) = f(L) = 0
Lap(1,1) = 0; Lap(1,2) = 0; Lap(2,1) = 0;    % So that f(0) = 0
Lap(N,N-1) = 0; Lap(N-1,N) = 0; Lap(N,N) = 0;% So that f(L) = 0

% To verify that D*f corresponds to taking the derivative of f
% and Lap*f corresponds to taking a second derviative of f,
% define f = sin(x) or choose your own f 
f = sin(x);
% And try the following:
Df = D*f; Lapf = Lap*f;
plot(x,f,'b',x,Df,'r', x,Lapf,'g');
axis([0 5 -1.1 1.1]);           % Optimized axis parameters

% To display the matrix D on screen, simply type D and return ...
D	% Displays the matrix D in the workspace
Lap 	% Displays the matrix Lap
\end{verbatim}
\end{program}

\section{Infinite Square Well}

When solving Eq.~(\ref{eq:one}), the method of separation of variables entails
that as an intermediate step we look for the separable solutions
\begin{equation}
|\Psi_E (t)\rangle =|\Psi_E (0)\rangle exp(-iEt/\hbar)
\label{eq:four}
\end{equation}
where $|\Psi_E (0)\rangle$ satisfies the time-independent
Schrodinger equation
\begin{equation}
{H} |\Psi_E(0) \rangle = E~|\Psi_E(0) \rangle.\label{eq:five}
\end{equation}
In solving Eq.~(\ref{eq:five}) we are solving for the eigenvalues $E$ 
and eigenvectors $|\Psi_E (0)\rangle$ of $H$. In MATLAB, the command 
$[V,E]=eig(H)$  does precisely this: it generates two
matrices. The first matrix  $V$ has as its columns the eigenvectors 
$|\Psi_E (0)\rangle$. The second matrix $E$ is a diagonal matrix with the
eigenvalues $E_i$ corresponding to the eigenvectors $|\Psi_{E_i} (0)\rangle$ 
placed along the central diagonal. We can use the command $E=diag(E)$ to 
convert this matrix into a column vector. In Program 3, we solve for 
the eigenfunctions and eigenvalues for the infinite square well Hamiltonian. 
For brevity, we omit the commands setting the parameters $L,N,x,$ and $dx$.

\begin{program}
\scriptsize
\begin{verbatim}
%****************************************************************
% Program 3: Matrix representation of differential operators,
% Solving for Eigenvectors & Eigenvalues of Infinite Square Well
%****************************************************************
% For brevity we omit the commands setting the parameters L, N, 
% x and dx; We also omit the commands defining the matrix Lap.
% These would be the same as in Program 2 above.

% Total Hamiltonian where hbar=1 and m=1
hbar = 1; m = 1; H = -(1/2)*(hbar^2/m)*Lap;
% Solve for eigenvector matrix V and eigenvalue matrix E of H
[V,E] = eig(H);

% Plot lowest 3 eigenfunctions
plot(x,V(:,3),'r',x,V(:,4),'b',x,V(:,5),'k'); shg;
E		% display eigenvalue matrix
diag(E)		% display a vector containing the eigenvalues
\end{verbatim}
\end{program}

Note that in the MATLAB syntax the object $V(:,3)$ specifies the
column vector consisting of all the elements in column 3 of matrix
$V$. Similarly $V(2,:)$  is the row vector consisting of all elements
in row 2 of $V$; $V(3,1)$ is the element at row 3, column 1 of $V$;
and $V(2,1:3)$ is a row vector consisting of elements $V(2,1)$,
$V(2,2)$ and $V(2,3)$.

\section{Arbitrary Potentials}

Numerical solution of Eq.~(\ref{eq:one}) is not limited to any 
particular potential. Program 4 gives example MATLAB codes solving 
the time independent Schrodinger equation for finite square
well potentials, the harmonic oscillator potential and even for
potentials that can only solved numerically such as the quartic
potential $U = x^4$.  In order to minimize the amount of RAM 
required, the codes shown make use of sparse matrices, 
where only the non-zero elements of the matrices are stored. The 
commands for sparse matrices are very similar to those for 
non-sparse matrices. For example, the command 
$[V,E]=eigs (H,nmodes..)$ provides the $nmodes$ lowest energy 
eigenvectors $V$ of of the sparse matrix $H$.

\begin{program}
\scriptsize
\begin{verbatim}
%****************************************************************
% Program 4: Find several lowest eigenmodes V(x) and 
% eigenenergies E of 1D Schrodinger equation
%         -1/2*hbar^2/m(d2/dx2)V(x) + U(x)V(x) = EV(x) 
% for arbitrary potentials U(x)
%****************************************************************
% Parameters for solving problem in the interval -L < x < L
% PARAMETERS:
L = 5;                      % Interval Length
N = 1000;                   % No of points
x = linspace(-L,L,N)';      % Coordinate vector
dx = x(2) - x(1);           % Coordinate step

% POTENTIAL, choose one or make your own
U = 1/2*100*x.^(2);    % quadratic harmonic oscillator potential
%U = 1/2*x.^(4);       % quartic potential

% Finite square well of width 2w and depth given
%w = L/50;
%U = -500*(heaviside(x+w)-heaviside(x-w));

% Two finite square wells of width 2w and distance 2a apart
%w = L/50; a=3*w;
%U = -200*(heaviside(x+w-a) - heaviside(x-w-a) ...
%              + heaviside(x+w+a) - heaviside(x-w+a));

% Three-point finite-difference representation of Laplacian 
% using sparse matrices, where you save memory by only 
% storing non-zero matrix elements
e = ones(N,1); Lap = spdiags([e -2*e e],[-1 0 1],N,N)/dx^2;

% Total Hamiltonian
hbar = 1; m = 1;      % constants for Hamiltonian
H = -1/2*(hbar^2/m)*Lap + spdiags(U,0,N,N);
	
% Find lowest nmodes eigenvectors and eigenvalues of sparse matrix
nmodes = 3; options.disp = 0;
[V,E] = eigs(H,nmodes,'sa',options);   % find eigs 
[E,ind] = sort(diag(E));% convert E to vector and sort low to high
V = V(:,ind);           % rearrange corresponding eigenvectors

% Generate plot of lowest energy eigenvectors V(x) and U(x) 
Usc = U*max(abs(V(:)))/max(abs(U));       % rescale U for plotting
plot(x,V,x,Usc,'--k');          % plot V(x) and rescaled U(x)
% Add legend showing Energy of plotted V(x) 
lgnd_str = [repmat('E = ',nmodes,1),num2str(E)]; 
legend(lgnd_str)                % place lengend string on plot 
shg
\end{verbatim}
\end{program}

\begin{figure}[ht]
\vskip 0.1 cm \includegraphics[angle=0,width=80mm]{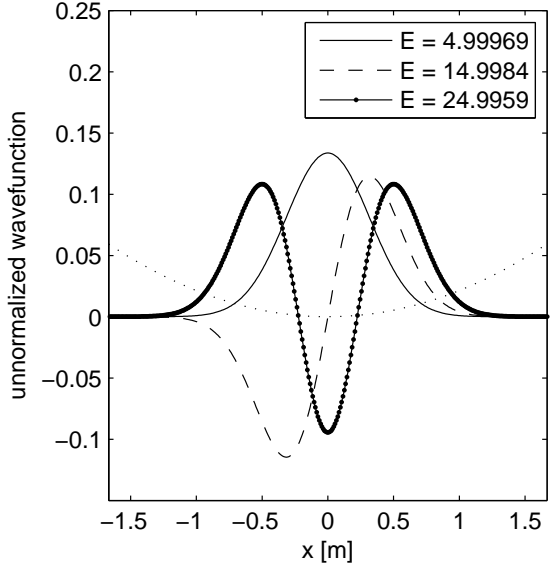} 
\vskip 0.1 cm \noindent
\caption{Output of Program 4, which plots the energy eigenfunctions 
$V(x)$ \textit and a scaled version of the potential 
$U(x) = 1/2 \cdot 100 \cdot x^{2}$. The corresponding energies displayed 
within the figure legend, 4.99969, 14.9984 and 24.9959, are, within
rounding error, precisely those expected from Eq.~(\ref{eq:harmonic}) for 
the three lowest-energy modes. 
\label{fig1} }
\end{figure}

Fig.~\ref{fig1} shows the plot obtained from Program 4 for the potential
$U= \frac{1}{2} \cdot 100 \cdot x^{2}$. Note that the 3 lowest energies
displayed in the figure are just as expected due to the analytic
formula
\begin{equation}
E = {\hbar \omega} \left( n + \frac {1}{2}\right) \label{eq:harmonic}
\end{equation}
with $n = \mbox{integer}$ and $\omega=\sqrt {\frac{k}{m}} = 10 $ rad/s.

\section{A Note on Units in our Programs}

When doing numerical calculations, it is important to minimize the 
effect of rounding errors by choosing units such that the variables 
used in the simulation are of the order of unity. In the programs 
presented here, our focus being undergraduate physics students, we 
wanted to avoid unnecessarily complicating matters. To make the equations
more familiar to the students, we explicitly left constants such as $\hbar$
in the formulas and chose units such that $\hbar =1$ and $m=1$. We 
recognize that others may have other opinions on how to address this
issue. An alternative approach used in research is to recast the equations 
in terms of dimensionless variables, for example by rescaling the 
energy to make it dimensionless by expressing it in terms of some 
characteristic energy in the problem. In a more advanced course for 
graduate students or in a course in numerical methods, such is an 
approach which would be preferable.

\section{Time Dependent Phenomena}

\begin{figure}[ht]
\begin{center}
\subfigure{ \includegraphics[scale=0.6]{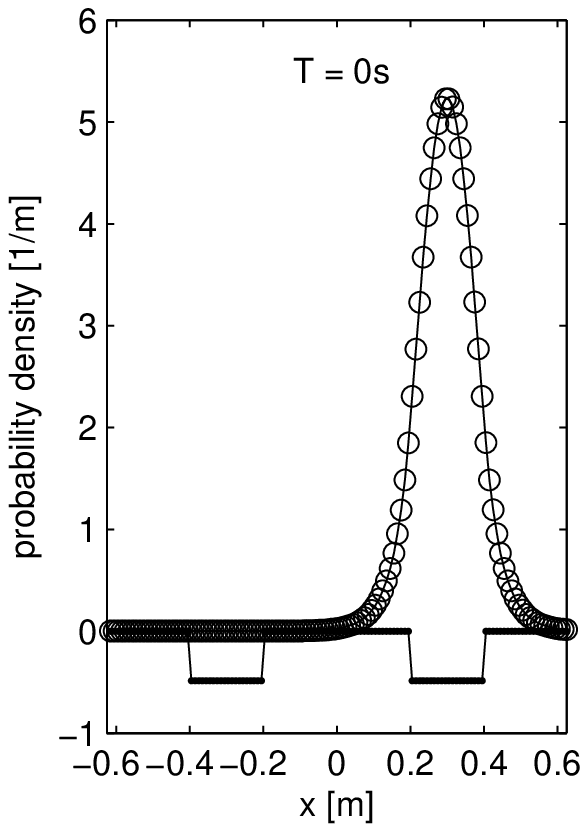}}
\subfigure{ \includegraphics[scale=0.6]{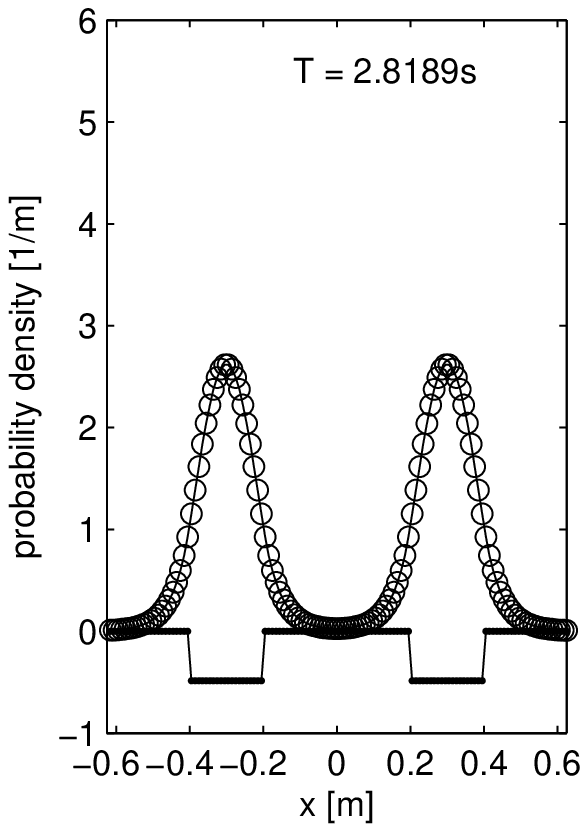}}
\subfigure{ \includegraphics[scale=0.6]{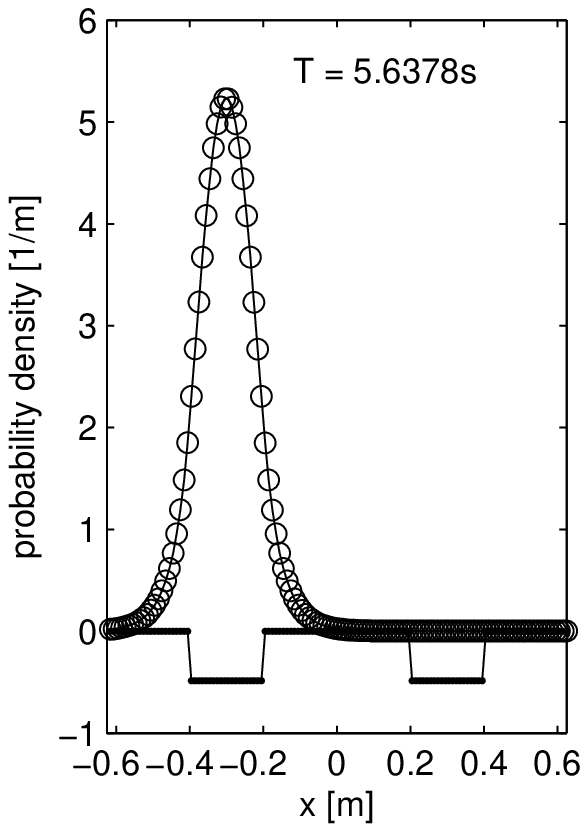}}
\subfigure{ \includegraphics[scale=0.6]{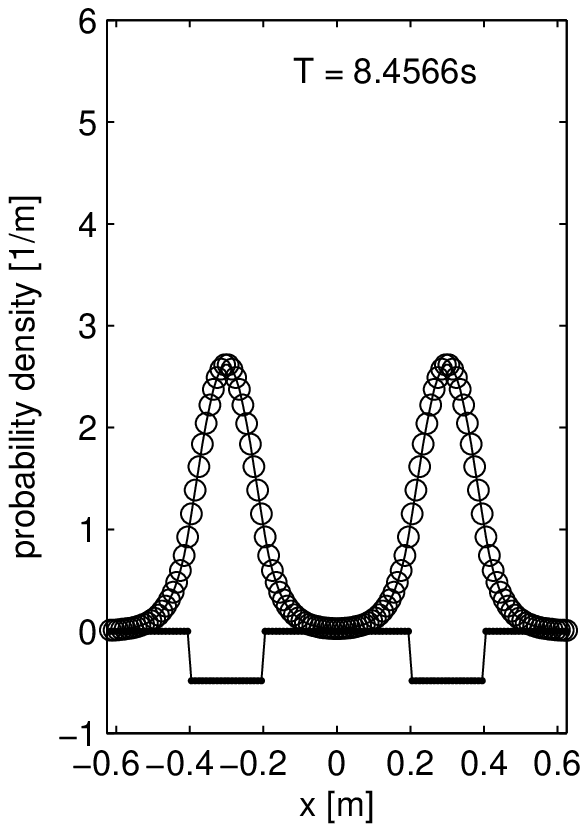}}
\subfigure{ \includegraphics[scale=0.6]{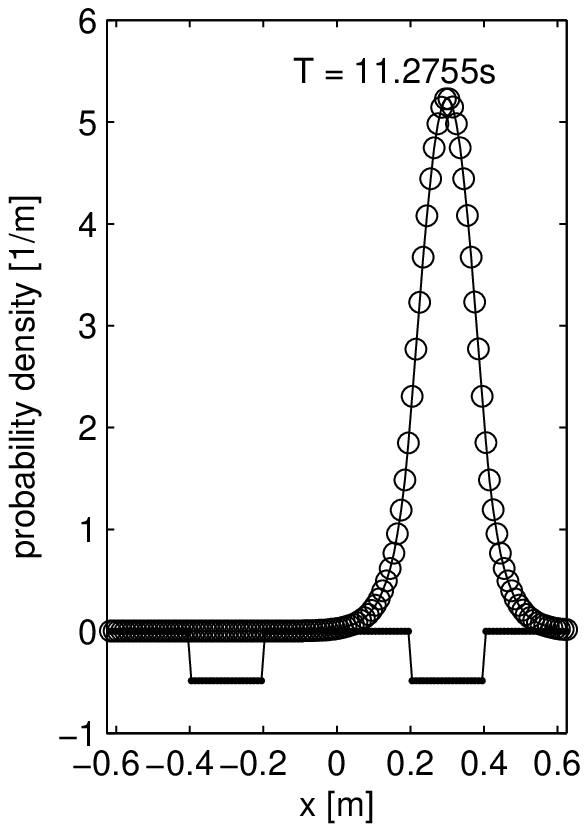}}
\vskip 0.1 cm \noindent
\caption{The probability density $\rho(x,t) $ output from Program 5 
for a particle trapped in a pair of finite-height square well potentials
that are closely adjacent. The initial state of the particle is chosen to be 
$|\Psi(0)\rangle \sim |E_0\rangle+|E_1\rangle$.
Shown is $\rho(x,t) $ plotted for times
$T=\lbrace 0,0.25,0.5, 0.75, 1.0\rbrace \times{2\pi \hbar}/{(E_1-E_2)}$. 
\label{fig2} }
\end{center}
\end{figure}

The separable solutions $|\Psi_E(t)\rangle $ are only a subset of 
all possible solutions of Eq.~(\ref{eq:one}). Fortunately, they are 
complete set so that we can construct the general solution 
\textit{via} the linear superposition
\begin{equation}
|\Psi (t)\rangle =\sum_E a_E |\Psi_E (0)\rangle exp(-iEt/\hbar)
\label{eq:six}
\end{equation} 
where $a_E$ are constants and the sum is over all possible values 
of $E$. The important difference between the separable solutions 
(\ref{eq:four}) and the general solution (\ref{eq:six}) is that the 
probability densities derived from the general solutions are 
time-dependent whereas those derived from the separable solutions 
are not. A very apt demonstration of this is provided in the 
Program 5 which calculates the time-dependent probability 
density $\rho(x,t) $ for a particle trapped in a a pair of 
finite-square wells whose initial state $|\Psi(0)\rangle $ is set 
equal to the the equally-weighted superposition 
\begin{equation} |\Psi(0)\rangle =\frac
{1}{\sqrt{2}}\left(|\Psi_{E_0}\rangle+|\Psi_{E_1}\rangle \right)
\end{equation}
of the ground state $|\Psi_{E_0}\rangle$ and first excited state 
$|\Psi_{E_1}\rangle$ of the double well system. As snapshots of the 
program output show in Fig.~\ref{fig2}, the particle is initially 
completely localized in the rightmost well. However, due to 
$E_0\neq E_1$, the probability density
\begin{eqnarray} 
\rho (x,t) = \frac {1}{2} [\; |\psi_{E_0}(x)|^2 
\; + \; |\psi_{E_1}(x)|^2   \; \;+ \; \nonumber \\
\;  2 |\psi_{E_0}(x)| |\psi_{E_1}(x)|
 cos^2\left({(E_1-E_2)t}/ {\hbar} \right)\; ] 
\label{eq:osc}
\end{eqnarray}
is time-dependent, oscillating between the $\rho(x)$ that corresponds 
to the particle being entirely in the right well
\begin{equation}
\rho (x) = \left|\; |\psi_{E_0}(x)|+ |\psi_{E_1}(x)|\; \right|^2
\end{equation}
and $\rho(x)$ for the particle being entirely in the left well
\begin{equation}
\rho (x) = \left|\; |\psi_{E_0}(x)|- |\psi_{E_1}(x)| \; \right|^2.
\end{equation}
By observing the period with which $\rho(x,t)$ oscillates in the 
simulation output shown in Fig.~\ref{fig2}
students can verify that it is the same as the period of oscillation 
${2\pi \hbar}/{(E_1-E_2)}$ expected from Eq.~(\ref{eq:osc}).

\begin{program}
\scriptsize
\begin{verbatim}
%****************************************************************
% Program 5: Calculate Probability Density as a function of time
% for a particle trapped in a double-well potential
%****************************************************************

% Potential due to two square wells of width 2w 
% and a distance 2a apart
w = L/50; a = 3*w;
U = -100*( heaviside(x+w-a) - heaviside(x-w-a) ...
                + heaviside(x+w+a) - heaviside(x-w+a));

% Finite-difference representation of Laplacian and Hamiltonian,
% where hbar = m = 1.
e = ones(N,1); Lap = spdiags([e -2*e e],[-1 0 1],N,N)/dx^2;
H = -(1/2)*Lap + spdiags(U,0,N,N);

% Find and sort lowest nmodes eigenvectors and eigenvalues of H
nmodes = 2; options.disp = 0; [V,E] = eigs(H,nmodes,'sa',options); 
[E,ind] = sort(diag(E));% convert E to vector and sort low to high
V = V(:,ind);           % rearrange coresponding eigenvectors
% Rescale eigenvectors so that they are always 
% positive at the center of the right well
   for c = 1:nmodes
       V(:,c) = V(:,c)/sign(V((3*N/4),c));
   end

%****************************************************************
% Compute and display normalized prob. density rho(x,T)
%****************************************************************
% Parameters for solving the problem in the interval 0 < T < TF
TF = 10;                  % Length of time interval
NT = 100;                 % No. of time points
T = linspace(0,TF,NT);    % Time vector

% Compute probability normalisation constant (at T=0)
  psi_o = 0.5*V(:,1)+0.5*V(:,2);    % wavefunction at T=0
  sq_norm = psi_o'*psi_o*dx;        % square norm = |<ff|ff>|^2

Usc = U*max(abs(V(:)))/max(abs(U));   % rescale U for plotting 

% Compute and display rho(x,T) for each time T
for t=1:NT;    % time index parameter for stepping through loop
  % Compute wavefunction psi(x,T) and rho(x,T) at T=T(t)
  psi = 0.5*V(:,1)*exp(-i*E(1)*T(t)) ...
            + 0.5*V(:,2)*exp(-i*E(2)*T(t));
  rho = conj(psi).*psi/sq_norm;   % normalized probability density

  % Plot rho(x,T) and rescaled potential energy Usc
  plot(x,rho,'o-k', x, Usc,'.-b'); axis([-L/8 L/8 -1 6]);
  lgnd_str = [repmat('T = ',1,1),num2str(T)]; 
  text(-0.12,5.5,lgnd_str, 'FontSize', 18);
  xlabel('x [m]', 'FontSize', 24); 
  ylabel('probability density [1/m]','FontSize', 24);
    
  pause(0.05);                    % wait 0.05 seconds
end
\end{verbatim}
\end{program}

\section{Wavepackets and Step Potentials}

Wavepackets are another time-dependent phenomenon encountered in
undergraduate quantum mechanics for which numerical solution techniques
have been typically advocated in the hopes of promoting intuitive 
acceptance and understanding of approximations necessarily invoked 
in more formal, analytic treatments. Program 6 calculates and displays 
the time evolution of a wavepacket for one of two possible potentials,
either $U=0$ or a step potential $U=U_o\Theta (x-L)$. The initial 
wavepacket is generated as the Fast Fourier Transform of a Gaussian 
momentum distribution centered on a particular value of the 
wavevector $k_o$. Because the wavepacket is composed of a distribution 
of different $k$s, different parts of the wavepacket move with 
different speeds, which leads to the wave packet spreading out in 
space as it moves.

While there is a distribution of velocities within the wavepacket, 
two velocities in particular are useful in characterizing it. The 
phase velocity $v_{w}=\omega/k=E/p=\hbar k_o/2m$ is the velocity 
of the plane wave component which has wavevector $k_o$. The group 
velocity $v_g=\hbar k_o/m$ is the velocity with which the expectation 
value $<x>$ moves and is the same as the classical particle velocity 
associated with the momentum $p=\hbar k$. Choosing $U=0$, students 
can modify this program to plot $<x>$ vs $t$. They can extract the 
group velocity from their numerical simulation and observe that 
indeed $v_g=2 v_w$ for a typical wave packet. Students can also 
observe that, while $v_g$ matches the particle speed from classical 
mechanics, the wavepacket spreads out as time elapses.

\begin{program}
\scriptsize
\begin{verbatim}
%****************************************************************
% Program 6: Wavepacket propagation using exponential of H
%****************************************************************
% Parameters for solving the problem in the interval 0 < x < L
L = 100;                        % Interval Length
N = 400;                        % No of points
x = linspace(0,L,N)';           % Coordinate vector
dx = x(2) - x(1);               % Coordinate step

% Parameters for making intial momentum space wavefunction phi(k)
ko = 2;                      % Peak momentum
a = 20;                      % Momentum width parameter
dk = 2*pi/L;                 % Momentum step
km=N*dk;                     % Momentum limit
k=linspace(0,+km,N)';        % Momentum vector

% Make psi(x,0) from Gaussian kspace wavefunction phi(k) using 
% fast fourier transform : 
phi = exp(-a*(k-ko).^2).*exp(-i*6*k.^2); % unnormalized phi(k)
psi = ifft(phi); % multiplies phi by expikx and integrates vs. x
psi = psi/sqrt(psi'*psi*dx);            % normalize the psi(x,0)

% Expectation value of energy; e.g. for the parameters 
% chosen above <E> = 2.062. 
avgE = phi'*0.5*diag(k.^2,0)*phi*dk/(phi'*phi*dk); 

% CHOOSE POTENTIAL U(X): Either U = 0 OR 
% U = step potential of height avgE that is located at x=L/2 
%U = 0*heaviside(x-(L/2));  % free particle wave packet evolution
U = avgE*heaviside(x-(L/2));    % scattering off step potential

% Finite-difference representation of Laplacian and Hamiltonian
e = ones(N,1); Lap = spdiags([e -2*e e],[-1 0 1],N,N)/dx^2;
H = -(1/2)*Lap + spdiags(U,0,N,N);

% Parameters for computing psi(x,T) at different times 0 < T < TF
NT = 200; 			% No. of time steps
TF = 29; T = linspace(0,TF,NT); % Time vector
dT = T(2)-T(1);			% Time step
hbar = 1;

% Time displacement operator E=exp(-iHdT/hbar)
E = expm(-i*full(H)*dT/hbar);	% time desplacement operator

%***************************************************************
% Simulate rho(x,T) and plot for each T
%***************************************************************
for t = 1:NT;         % time index for loop
    % calculate probability density rho(x,T)
    psi = E*psi;            % calculate new psi from old psi
    rho = conj(psi).*psi;   % rho(x,T)

    plot(x,rho,'k');     % plot rho(x,T) vs. x
    axis([0 L 0 0.15]);  % set x,y axis parameters for plotting
    xlabel('x [m]', 'FontSize', 24); 
    ylabel('probability density [1/m]','FontSize', 24);

    pause(0.05);         % pause between each frame displayed
end

% Calculate Reflection probability
R=0;
for a=1:N/2;
    R=R+rho(a);
end
R=R*dx
\end{verbatim}
\end{program}

In Program 6, we propagate the wave function forward 
\textit{via} the formal solution
\begin{equation} |\Psi(t)\rangle = exp(-iHt/\hbar)|\Psi(0)\rangle,
\label{eq:seven}
\end{equation}
where the Hamiltonian matrix $H$ is in the exponential. This solution
is equivalent to Eq.~(\ref{eq:five}), as as can be shown by simple 
substitution. Moreover, MATLAB has no trouble exponentiating 
the matrix that numerically representing the Hamiltonian operator 
as long as the matrix is small enough to fit in the available 
computer memory.

\begin{figure}[ht]
\subfigure{ \includegraphics[scale=0.4]{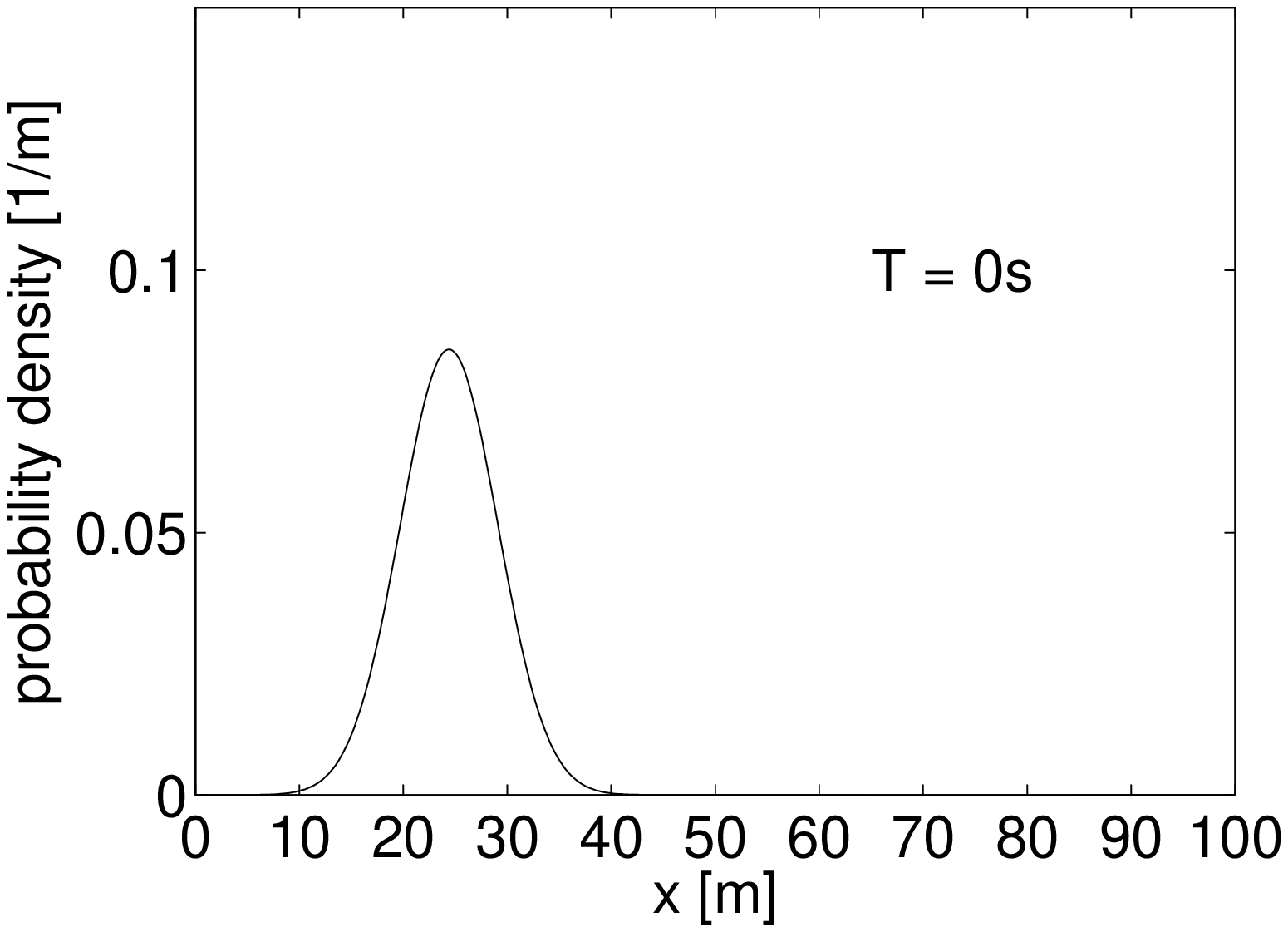}}
\subfigure{ \includegraphics[scale=0.4]{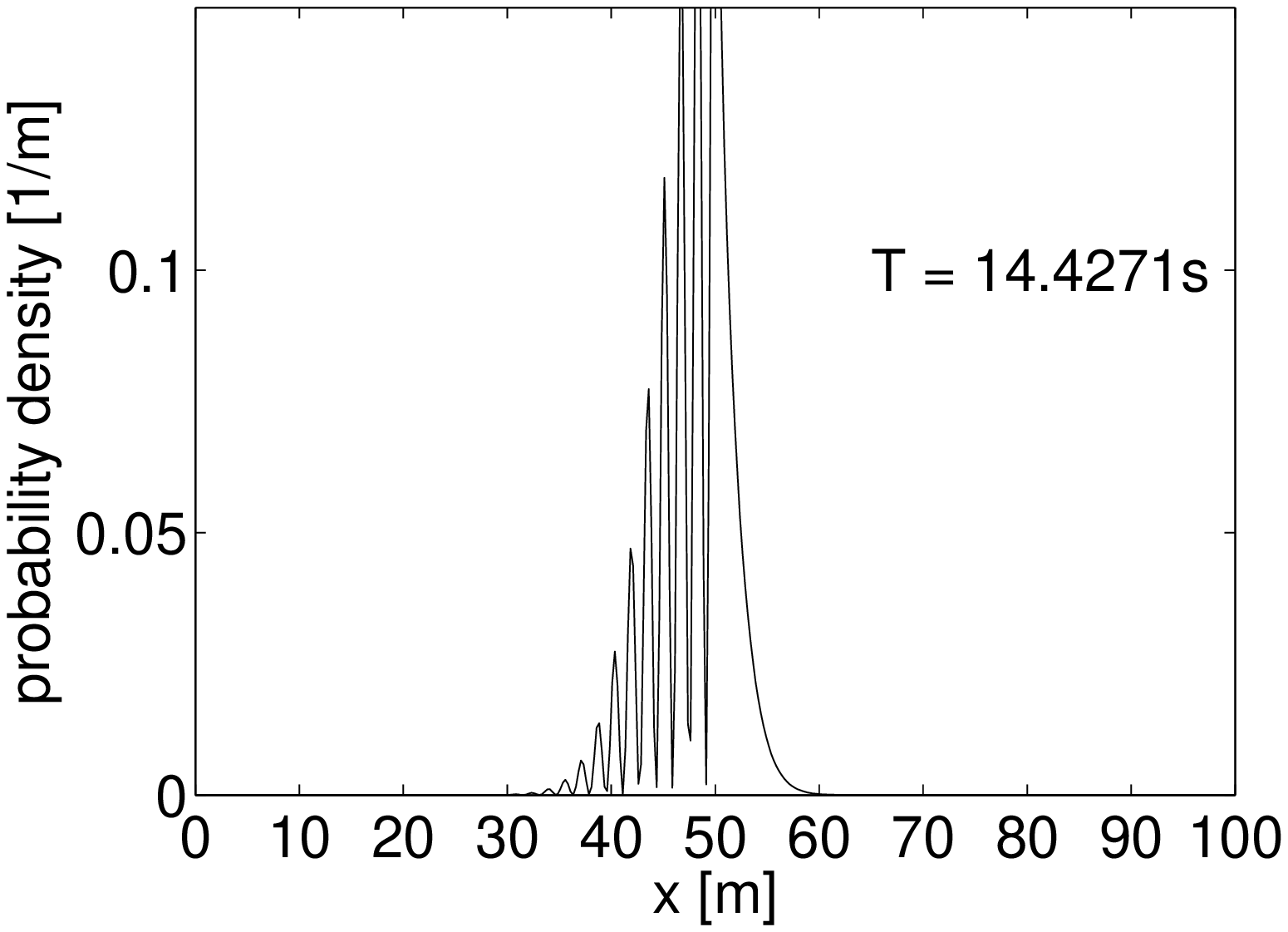}}
\subfigure{ \includegraphics[scale=0.4]{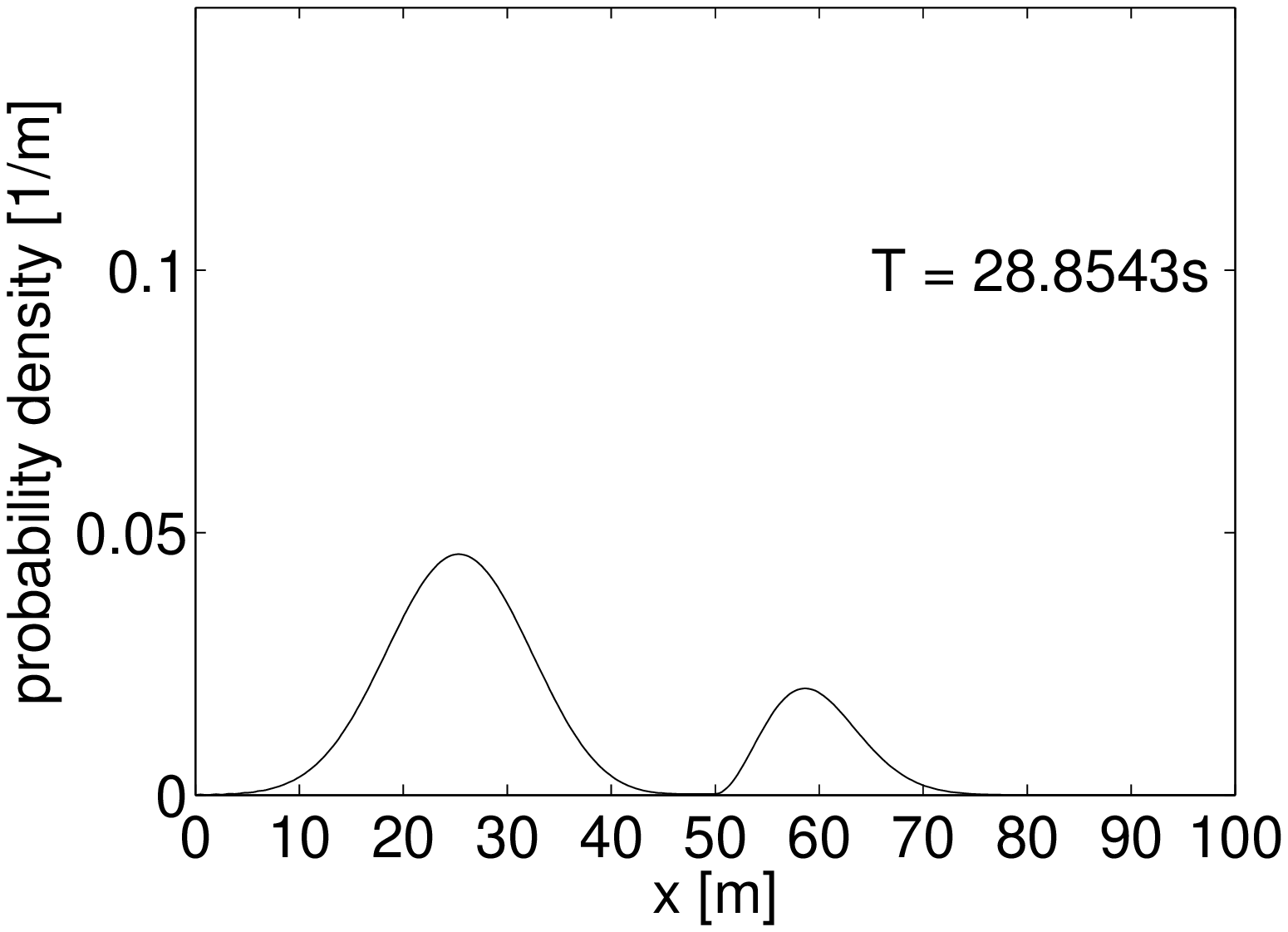}}
\vskip 0.05 cm \noindent
\caption{Output of Program 6 showing a wavepacket encountering 
step potential of height $\sim \langle E\rangle $ located 
at $x/L=0.5$ at different times. \label{fig3} }
\end{figure}

Even more interestingly, students can use this method to
investigate scattering of wavepackets from various potentials, 
including the step potential $U=U_o \Theta(x-L/2)$.
In Fig.~\ref{fig3}, we show the results of what happens as the wavepacket
impinges on the potential barrier. The parameters characterizing 
the initial wavepacket have been deliberately chosen so that the wings 
do not fall outside the simulation area and initially also do not 
overlap the barrier on the right. If $\langle E\rangle \ll U_o$, 
the wavepacket is completely reflected from the barrier. 
If $\langle E\rangle \approx U_o$, a portion of the wave is 
is reflected and a portion is transmitted through. If 
$\langle E\rangle ~\gg U_o$, almost all of the wave is 
transmitted.

\begin{figure}[ht]
\vskip 0.1 cm 
\includegraphics[angle=0,width=85mm]{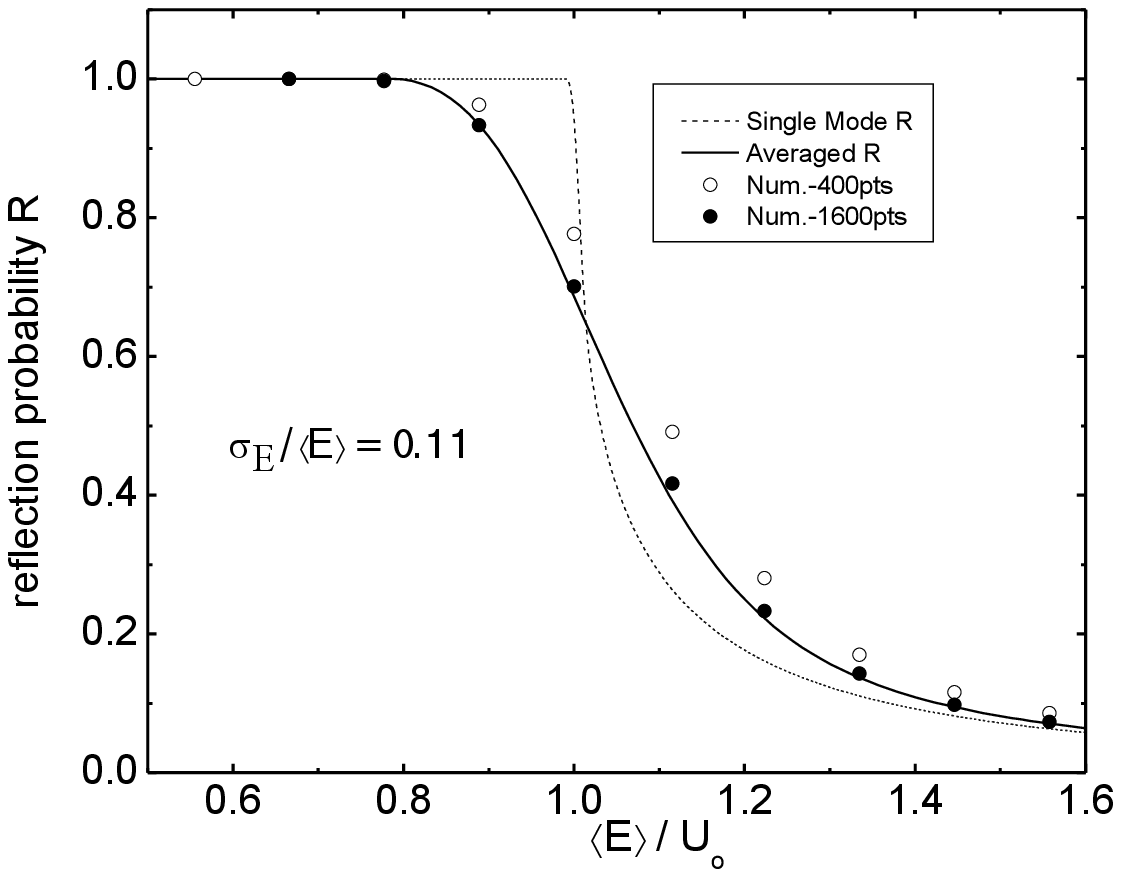} 
\vskip 0.1 cm \noindent
\caption{The reflection probability $R$ \textit{vs.} 
$\langle E\rangle/U_o$. The dashed line 
is simply Eq.~(\ref{eq:eight}) where we subsitute 
$E=\langle E \rangle$, the solid line is 
Eq.~(\ref{eq:eight}) averaged over the energy distribution
in the incident wavepacket, and the points are numerical 
results obtained using Program 6, where the horizontal 
distance between points is $\sigma_E / U_o$ where
$\sigma_E$ is the standard deviation of the energy 
distribution in the wavepacket. \label{fig4} }
\end{figure}

In Fig.~\ref{fig4} we compare the reflection probability $R$ 
calculated numerically using Program 6 with $R$ calculated 
by averaging the single-mode \cite{griffiths} expression 
\begin{equation}
R(E) = \left|\frac {\sqrt{E}-\sqrt{E-U_o}}{\sqrt{E}+\sqrt{E-U_o}}
\right|^2 \label{eq:eight}
\end{equation}
over the distribution of energies in the initial wavepacket.
While the numerically and analytically estimated $R$ are found to
agree for large and small $\langle E\rangle / U_o$, there is a
noticeable discrepancy due to the shortcomings of the numerical 
simulation for $\langle E\rangle /U_o \approx 1$. 
This discrepancy can be reduced significantly by increasing the 
number of points in the simulation to 1600 but only at the cost of
significantly slowing down the speed of the computation. For our
purposes, the importance comparing the analytical and numerical
calculations is that it gives student a baseline from which to 
form an opinion or intuition regarding the accuracy of 
Eq.~(\ref{eq:eight}).

\section{Conclusions}

One benefit of incorporating numerical simulation into the teaching
of quantum mechanics, as we have mentioned, is the development of
student intuition. Another is showing students that non-ideal,
real-world problems can be solved using the concepts they learn
in the classroom. However, our experimentation incorporating these 
simulations in quantum physics at WPI during the past year has shown 
us that the most important benefit is a type of side-effect to doing 
numerical simulation: the acceptance on an intuitive level by the 
student that functions are like vectors and differential operators 
are like matrices. While in the present paper, we have only had
sufficient space to present the most illustrative MATLAB codes, 
our goal is to eventualy make available a more complete set of 
polished codes is available for downloading either from the
authors or directly from the file exchange at MATLAB Central.\cite{matlabcentral}



\begin{thebibliography}{99}
\bibitem{proportionality} Born's law is stated as a proportionality because
an additional factor is necessary depending on the units of $|\Psi \rangle$
\end{thebibliography}
\end{document}